\documentclass[12pt,a4paper ]{article}
\usepackage[utf8]{inputenc} 
\usepackage{amsmath,amsfonts,amssymb,euscript}
\usepackage[english, russian]{babel}
\date{}
\title{{\large {Solution of a System of Linear Equations in an Integral Ring 
\thanks{
USSR J. of Comput. Math. and Math. Phys., V.23, No. 6, 1983. 1497-1500.  (Engl.) 
Малашонок Г.И. Решение системы линейных уравнений в целостном кольце.  
Журнал вычислительной математики и математической физики. 1983. Т.23. No.6. 1497–1500.(Russian)}
}}}
\author{
G.I. Malaschonok \\
(Lviv, Ukraine)}
\begin{document}
\maketitle

\begin{abstract}
A modified Gauss's algorithm for solving a system of linear equations in an integral ring is proposed, 
as well as an appropriate algorithm for calculating the elements of the adjoint matrix.
\end{abstract}

\section{Introduction}
Consider the system of linear equations
$$
\sum_{j=1}^n a_{i,j}x_j = a_{i,n+1}, \ \ \ \  i=l,2,\ldots , п, \eqno(1.1)
$$
whose coefficients and constants belong to the integral commutative 
ring R.We shall express this system using the augmented $n\times (n+l)$ matrix:
$$
  B=\|a_{ij}\|,  i=l,2,\ldots ,n, j=1,2,\ldots ,n+l.\eqno(1.2)
$$
As usual, we will introduce a matrix formed by the coefficients 
of system (1.1) 
and denote it by $A$, and also the matrices $A_i$, $i=1,2,\ldots,n$,
which are obtained from matrix $A$ by replacing the elements of 
the i-th column by the constant terms.
 We will denote the determinants of $A$ and $A_i$ by $\Delta$ and $\Delta_i$, respectively. 

 Generally speaking, the solution of (1.1) belongs to the field of 
 quotients of the ring $R$ (see. [1], [2]). 
However, there are algorithms for solving system (1.1) 
whose intermediate results are in the ring $R$. 
Examples of such algorithms are the scheme of multiplication and 
subtraction of the Gauss method [3], or the method of solution 
using the Cramer formulas
$$
  x_i=\Delta_i/ \Delta,   i=l,2,\ldots,п,\eqno(1.3)
$$
when the determinants are found by expansion with respect to rows 
and columns.

Such algorithms of solution in a ring have the advantage of avoiding 
the intermediate rounding-off errors; they also simplify the 
operations. However, a solution using Cramer’s formulas with the 
immediate finding of determinants by the operations of summation, 
subtraction and multiplication only, requires a vast amount of 
calculation; the multiplication and subtraction of 
the Gauss method results in a rapid increase in the number of digits.
 
The proposed algorithm for the solution in an integral commutative 
ring is free from these drawbacks, since it uses the reduction of 
both sides of an equation to a non-zero common multiplier: this is 
permissible in an integral ring.

As usual, we first perform a forward way and reduce the matrix $B$
to the equivalent triangular matrix $B_n$ and then a backward way to 
arrive at the diagonal matrix $D_1$.

\section{Transformation to the triangular form}
 We denote by $A_{ij}^k$,  $l\leq k\leq n$, $ k\leq j\leq n+1$, 
 a square matrix of order $k$, which represents the left-hand top part 
 of matrix $B$ bordered by row i and column j, that is
$$
A_{ij}^k=\|a_{pq}\|, \ \  p=1,2,\ldots,k-1, \ \  i, q=l,2,\ldots,k-1,j,
\eqno(2.1)
$$
and introduce the symbol for its determinant:
$$
a_{ij}^k=|A_{ij}^k|.  \eqno(2.2)
$$
We note that $a_{ij}^1=a_{ij}, \  i=1,2,\ldots,n, \  j=l,2,\ldots,n+l$.

We assume that the condition for the realization of the Gauss 
algorithm is satisfied, that is the diagonal minors are 
non-zero:
$$
   a_{kk}^k \neq 0,   k=i,2,\ldots,n. \eqno(2.3)
$$
At the first step we transform matrix $B$ using the scheme of 
multiplication and subtraction [3]: from each row with index $i$, $i\geq 2$, 
multiplied by $a_{11}$ we subtract the first row multiplied by $a_{11}$.
As a result we have the matrix
$$
B_2=\|b_{i,j}^2\|, \ \ b_{1j}^2=a_{1j}^1, \ \ b_{i1}^2=0, \ \  b_{ij}^2=a_{ij}^2, \ \ i>1,
$$
whose elements are determined by formulas (2.2).

Each following step, with the exception of the first, 
requires reduction to the common multiplier. The  
$k$-th step, $k\geq 2$, is performed as follows; at the 
preceding step let the 
following matrix with the element (2.2) be obtained:
$$
B_k=\|b_{ij}^k\|,  \ \ \ \ b_{ij}^k= 
\left\{\begin{array}{ccl}
a_{ij}^i & \hbox{ \ for \ } & i=1,2,\ldots,k-1, \ j\geq i, \\
0        & \hbox{ \ for \ } & i>j, j=1,2,\ldots,k-1,  \\
a_{ij}^k & \hbox{ \ for \ } & i, j\geq k.
\end{array}
\right.
\eqno(2.4)
$$
We transform the matrix $B^k$ at the $k$-th step a follows: 
from each row with index $i$,  $i\geq k+1$, multiplied by $a_{kk}^k$, 
we subtract the row with index $k$, multiplied by $a_{ik}^k$, then 
the element of the $j$-th column, $j\geq k+1$, and the row $i$ take 
the form 
$$
  a_{kk}^k a_{ij}^k-a_{ik}^k a_{kj}^k \eqno (2.5)
$$
and in the first columns will be zeros only.

The distinctive feature of this method is that all the 
elements (2.5) can be reduced to
the leading element $a_{k-1,k-1}^{k-1}$ and the matrix $B_{k+1}$
can be obtained in accordance with  (2.4).

To demonstrate this, we make use of the Sylvester determinant 
identity [4], which for matrix $A$ of order $n$ and for any $s$,
$s=l,2,\ldots,n-1$, can be written in the adopted notatione 
(2.1) and (2.2) as
$$
|M^s|=(a_{s-1,s-1}^{s-1})^{n-s} \Delta,   \eqno(2.6)
$$
where $M^s$, $s=l,2,\ldots,n$, 
are matrices of order $n-s+1$ with elements	
 $a_{pq}^s$, $p, q = s, s+l,\ldots,n$.

We note that the Sylvester determinant identity is valid for 
matrices over a commutative integral ring; to show this, it is
enough to make use of Gauss’s method using the scheme of 
multiplication and subtraction.

We write the Sylvester determinant identity for matrix
$A_{i,j}^{k+1}$ of order $k+l$, $k<i\leq n$, $k<j\leq n+1$, 
with the condition $s=k$:
$$
\left| \begin{array} {cc}
a_{kk}^k & a_{kj}^k \\
a_
{ik}^k  & a_{ij}^k \end{array}
\right| = a_{k-1,k-1}^{k-1}   a_{ij}^{k+1}.
$$
This identity shows that expression (2.5) can be decomposed into a 
product of two cofactors, and thus it is possible to reduce all rows, 
starting with the (k+1)-th, to the common multiplier$a_{k-1,k_1}^{k-1}$,
which in accordence with (2.3) is non-zero, and as a rasult
we can obtain matrix $B_{k+1}$.

After the $(n-1)$-th step of the forward way we obtain the 
desirad triangular matrix $B_n$.
In it, on the principal diagonal are the diagonal minors 
$a_{kk}^k$, 
$k=1,2,\ldots,n$, and all the elemante under it are zeros.

\section{Transformation to the diagonal form}
We denote the $n\times (n+1)$ matrix as follows:
$$
 D_k= \|d_{ij}\|, \ \ \ k=1,2,\ldots,n, \eqno (3.1a)
$$
$$
d_{i,j}= 
\left\{\begin{array}{ccl}
0        & \hbox{ \ при \ } & i>j, \hbox {\ и при \ } i=k,\ldots,n-1, \ j=i+1,\ldots,n;  \\
a_{ij}^i & \hbox{ \ при \ } & i=1,2,\ldots,k-1, \ j\geq i;
\end{array}
\right.
\eqno (3.1b)
$$
$$d_{ii}=\Delta; d_{i,n+1} = \Delta_i, \  i=k,\ldots,n.
 \eqno (3.1c)
$$

Note that it followa from definition (2.2) that $a_{nn}^n=\Delta$ and  
$a_{n,n+1}^n=\Delta_n$. Therefore $D_n=B_n$. 

Thus, at the first step of the backward way we already have matrix $D_n$.

Let the matrix $D_k$, $k=2,3,\ldots,n$ be obtained at a certain step 
of this run. To perform the next step we multiply all elements of 
row $k-1$ by $\Delta$ and subtract from it all the lower 
rows with indices $i$, $i=k, k+i,\ldots,n$, 
previously multiplying them by $a_{k-1,i}^{k-1}$. 
As a  result, in the (k-1)-th row the diagonal element will be 
$a_{k-1,k-1}^{k-1}\Delta$, the elements in the last 
column will become 
$$
a_{k-1,n+1}^{k-1}\Delta- \sum_{i=k}^{n} a_{k-1,i}^{k-1}\Delta_i, \eqno(3.2)
$$
and all the remaining elements become zero elements.

We will now show that expression (3.2) can bs broken down into ths 
multipliers  $\Delta_{k-1}$ and
$a_{k-1,k-1}^{k-1}$. To do this we introduce square matrices $M_j^s$, 
$1\leq s\leq j\leq n$,  formed by replacing in the matrix $M^s$
the $j$-th column by the $(n+1)$-th; then we write the Sylvester 
identity for 
matrices $A_j$, and for any $s$, $s=1,2,\ldots,n$, in the form
$$
|M_j^s|= (a_{k-1,k-1}^{k-1})^{n-s}\Delta_j. \eqno(3.3)
$$
We substitute $|M_j^{k-1}|$ and $|M^{k-1}|$ from Eqs. (2.6) and 
(3.3) into the identity 
$$
\sum_{j=k-1}^{n} a_{k-1,j}^{k-1}|M_j^{k-1}|=a_{k-1,n+1}^{k-1} |M^{k-1}|,
$$
which can be easily proved by expanding it with respect to column j, 
and changing the order of summation on the left-hand side.

After reducing $(a_{k-2,k-2}^{k-2})^{n-k+1}$ and regrouping terms, 
we arrive at the identity
$$
a_{k-1,n+1}^{k-1}\Delta- \sum_{i=k}^{n} a_{k-1,i}^{k-1}\Delta_i=
a_{k-1,j}^{k-1}  \Delta_{k-1}.
$$
Thus, we have shown that the elements of the $(k-1)$-th row can be 
reduced to $a_{k-1,k-1}^{k-1}$, and as a result we obtain matrix
$D_{k-1}$ in 
conformity with (3.1).

After $n-1$ steps of the backward way we arrive at the diagonal matrix 
$D_1$ where all elements on the principal diagonal equal 
the determinant $\Delta$,
of the basic matrix $A$; in the last column in the $i$-th row,	
 $i=l,2,\ldots,n$, we have determinant $\Delta_i$, 
 and all the remaining elements
are zeros.
 
Thus, a solution using Cramer's formulas (1.3) has been obtained; 
all the determinants $\Delta$ and $\Delta_i$  have been evaluated 
simultaneously, all the time inside the ring $R$.

\section {The algorithm}
We will now write the algorithm separately, 
assuming that the condition of feasibility of Gauss's algorithm 
(2.3) is satisfied.
 
{\b Forward way}.

1. We multiply each row of the augmented matrix (1.2), with 
$i$, $i\geq 2$ by $a_{11}$ and subtract from it the first row 
multiplied by $a_{i1}$. Thus we arrive at the matrix $B_2$.
 
  2. Let the matrix $B_k$, $k=2,3,\ldots,n-1$, be obtained at the 
  preceding step. We multiply each row with number $i$, $i\geq k+1$ by
  $a_{kk}^k$ and subtract from it the row with number $k$, multiplied 
by $a_{ik}^k$.
The rows with numbers $k+1, k+2,\ldots, n$  are reduced to 
$a_{k-1,k-1}^{k-1}$ . 
Thus we obtain the matrix $B_{k+1}$.
 
3. After the $(n-1)$-th step we arrive at the triangular matrix $B_n$,  
in which the outermost row has two non-zero elements $a_{nn}^n=\Delta$ and $a_{n,n-1}^n=\Delta_n$.

{\bf Backward way}. 

1. Let $\Delta_n, \Delta_{n-1}, \ldots, \Delta_{k+1}$, $1\leq k \leq n-1$	
be already known; then, using the elements of
the matrix $B_n$ we find $\Delta_k$ from the formula
 $$
\Delta_k={ \Delta a_{k,n+1}^{k}- \sum_{j=k+1}^{n} a_{k,j}^{k}\Delta_j
\over a_{kk}^k},
$$
where the fraction stroke means that division without a remainder 
is possible.

2. The solution of the system is found using Cramer's formula
$x_i =\Delta_i/ \Delta, i= 1,2,\ldots,n$.

\section{Calculation of the elements of the adjoint matrix}
The calculation of the adjoint matrix in an integral commutative 
ring with unit element is an important application of the 
method discussed.
 
Let $P=\|p_{ij}\|$, $i,j=1,2,\ldots n$, 
-- be an adjoint of the matrix
$A =\|a_i,j\|$, $i,j=1, 2,\ldots ,n$;
 then finding its elements reduces [3] to solving $n$ systems of the 
 form   
$$
\sum_{k=1}^n a_{ik} p_{kj} = \Delta \delta_{i,j}, \ \ \ 
  i=l,2,\ldots,n, \ \ \  j=1,2,\ldots, n, 
$$
where $\delta_{i,j}$ is the Kronecker delta, and $\Delta=|A|$.

The forward way for all $n$ matrices can be performed simultaneously 
if we consider the matrix
$$
B_E=\left\| \begin{array}{ccccccc}
a_{11} & \ldots & a_{1n} & 1 & 0 & \ldots & 0 \\
a_{21} & \ldots & a_{2n} & 0 & 1 & \ldots & 0 \\
\ldots & \ldots & \ldots & \ldots & \ldots & \ldots & \ldots \\
a_{n1} & \ldots & a_{nn} & 0 & 0 & \ldots & 1 \\
\end{array}
\right\|.
$$
instead of (1.2).

As a result we obtain the matrix 
$B_{En}=\|a_{ij}^n\|$, $i=1,2,\ldots,n$, $j=1,2,\ldots,2n$, 
in which all the elements below the principal diagonal and above the 
outermost diagonal are zeros, and the non-zero elements in the last 
row are
$a_{nn}^n=\Delta$ and $a_{n,n+j}^n=p_{nj}$, $j=1,2,\ldots,n$.

The backward way is perfqrmed in the usual way: let the elements 
$p_{ij}$,  $i=k+l, k+2,\ldots,n$, $j=1,2,\ldots,n$, $1\leq k \leq n-1$,
  of the last $n-k$ rows be known; then the elements of the 
$k$-th row of the adjoint matrix are obtained, using the elements of 
$B_{En}$, by the formula
$$
p_{kj}={ \Delta a_{k,n+j}^{k}- \sum_{i=k+1}^{n} a_{k,i}^{k} p_{ij}
\over a_{kk}^k},  \ \ \ j=1,2,\ldots,n,
$$
where the fraction line indicates division without a remainder.

It is obvious that the necessary amount of operations for the above 
algorithm to solve a system of equations, and for the corresponding 
algorithm for finding the elements of an adjoint matrix is of the
same order as that for Gauss's method.

\end{document}